# Collective Transport of Magnetic Microparticles at a Fluid Interface through Dynamic Self-Assembled Lattices

*Fernando Martínez-Pedrero[1,\*], Francisco Ortega[1,2], Ramón G.Rubio[1,2], Carles Calero[3,4,\*]*

Dr. F. Martínez-Pedrero, Prof. F. Ortega, Prof. R. G. Rubio
Departamento de Química-Física, Universidad Complutense de Madrid, Avda. Complutense s/n, Madrid, 28040, Spain
E-mail: fernandm@ucm.es

Prof. F. Ortega, Prof. R. G. Rubio
Inst. Pluridisciplinar, Universidad Complutense de Madrid, Paseo Juan 23,1, E-28040 Madrid, Spain

Dr. C. Calero
Condensed Matter Physics Department, Faculty of Physics, Universitat de Barcelona, Spain
Institut de Nanociència i Nanotecnologia, IN2UB, Universitat de Barcelona, Barcelona, Spain
E-mail: carles.calero@ub.edu

The transport of motile entities across modulated energy landscapes plays an important role in a range of phenomena in biology, colloidal science and solid-state physics. Here, an easily implementable strategy that allows for the collective and monitored transport of microparticles at fluid-fluid interfaces is introduced. Adsorbed magnetic microparticles are carried on time-dependent magnetic potentials, generated by a dynamic self-assembled lattice of different-sized magnetic particles. In such binary systems, the sudden reorientation of the applied field triggers the rapid exchange between attractive and repulsive configurations, enabling for the ballistic transfer of the carriers through the lattice. As the number of motile entities increases, the induced current increases, before reaching a maximum, while the loaded interface gradually displays bidirectional transport. The described methodology can be tuned through the applied field and exploited for the monitored guidance of adsorbed molecules on liquid surfaces, the segregation of colloidal mixtures, the induced motion of defects in photonic crystals or the design of new self-assembled microrobots.

# 1. Introduction

The controlled transport of particles through ordered or spatially random lattices is one of the main challenges in many areas of surface science and biology.[1] In those systems, the velocity and direction of the mobile entities, which travel on steady or time-dependent potentials frequently generated on plane or curved surfaces, are determined by the lattice geometry, the driving mechanism, and the emergence of transport anomalies and collective phenomena.[2] These processes occur in a rich variety of systems: adatoms adsorbed on a crystal surface,[3] atoms on the surface of metals,[4] metal nanocrystals on multiwall carbon nanotubes,[5] electrons in graphene layers,[6] membrane proteins and lipids[7], or colloids in micro-structured substrates[2, 8-10] and at fluid-fluid interfaces.[11]

For particles adsorbed at fluid interfaces, ubiquitously found in industrial products and used as model platforms for the design of advanced materials, the methods of transport designed so far exploit interfacial physical effects, via surface waves,[12, 13] magnetocapillary and Marangoni effects,[14, 15] self-generated gradients,[11] or the rectification of the hydrodynamic flows generated by rolling species.[16-18] The proposed strategies, strongly determined not only by the in-plane spatial distribution of the adsorbed species but also by small variations in the wettability,[19-21] are able to overcome the restriction imposed by the scallop theorem, but not to generate a colloidal current.

The collective behavior of the transported particles under confinement is found to be relevant in a broad range of systems.[8, 11, 17, 22] Confined micro-swarms are found in a broad range of synthetic and biological systems, such as in the cooperative transport of molecular motors,[23] colloidal rollers on liquid/solid interfaces,[24, 25] polarizable microparticles enclosed in narrow volumes,[26, 27] growing biofilms[28] or motile synthetic designs and microorganisms confined in microfluidic channels, ducts, blood vessels or fluid and solid interfaces.[29-35] They exhibit different kind of collective behaviors,[36] ranging from alignment,[25] flocking,[12, 36-38] glassy transitions,[39] jamming and clogging phenomena, with temporary or permanent arrest and crystallization.[40-45]

In this work we demonstrate for the first time, both experimentally and with numerical simulations, a new mechanism of transport, that allows for the collective and monitored transport at the microscale of particles strongly confined to a fluid interface. The transport mechanism is based on the use of two types of magnetic particles that adsorb at different vertical positions of the interface, which in our case is achieved by using particles of different sizes with affinity to one of the fluids. The vertical distance between particles and the anisotropy of the dipole-dipole interaction allows the existence of states where the induced dipoles at a given plane repel one another whereas dipoles at different planes attract forming hetero-dimers, oriented along the horizontal component of the tilted applied field. The sudden reorientation of the field can turn the formed heterodimers into unstable configurations, producing the rapid mutual ejection of the particles and the subsequent creation of new heterodimers oriented along the new direction of the field. This process occurs while the interaction between particles confined at the same plane remains repulsive. In loaded interfaces with sufficiently high colloidal surface density, this principle can be used to induce the steered transport of one of the two species through the 2D self-assembled lattice formed by their counterparts. This is done by modulating the dipole-dipole interactions through the actuation with an external time-dependent magnetic field, which precesses about the axis perpendicular to the interface and recursively inverts its vertical component.

We also combined experiment and numerical simulations to analyze the dependence of the transport properties on the features of the actuating field. We find that the dynamics of the transferred particles presents a crossover from a diffusive to a ballistic regime, determined by the surface density, the angle $\gamma$ subtended by the applied field with the interface, the precessing frequency and its intensity. In the ballistic regime, the trajectories are determined by the time pattern of inversion of the component perpendicular to the fluid interface. If this component is inverted at a frequency double the precessing frequency, the moving particles are transported along a crystallographic direction through the lattice. If the inversion of the field is controlled on the fly, the particles can be steered in any 2D-planned trajectory. Finally, we quantify the effect of concentration of moving particles on their transport. We find that the gradual increase in the number of motile entities leads first to the increase in the current, then to the appearance of a maximum flow, and finally to an inverted scenario, where the particles that initially formed the lattice are transported on a monolayer of the previous motile ones.

## 2. Results

### 2.1 Transport Mechanism

*2.1.1. Transference of Different-Sized Particles in Diluted Monolayers*

In a binary system of paramagnetic particles of different size, with radius $R_B$ and $R_S$, adsorbed onto a fluid interface, the particle centers are confined to planes separated by a distance $\Delta z$, In our case, and according to the SEM images of the magnetic particles obtained with the Gel Trapping Technique (GET) (see inset in Figure 1a and Reference [20]), the affinity of the two types of particles is preferentially to the water phase, and $\Delta z \approx R_B - R_S = 0.9$ μm. The application of a field in the X direction parallel to the interface, $H_x$, induces the formation of linear aggregates, composed by small and big particles aligned along the X axis. The subsequent application of an extra field perpendicular to the interface, $H_z$, modulates the magnetic dipole-dipole interaction between colloidal particles, promoting the repulsion between the particles, the fragmentation of the linear aggregates and the generation of a set of new states, determined by the field strength $H_0 = \sqrt{H_x^2 + H_z^2}$ and tilt angle $\gamma = a\tan(H_z/H_x)$ (see **Figure S1** in Supporting Information (SI) and Reference [20]). In the range 54,7º < γ < (asin (Δz/($R_B$ + $R_s$)) + 54.7º), equally sized magnetic particles repel one another, whereas pairs of big and small particles, can attract and form heterodimers. The formed dimers are directed along the X axis and are ordered such that the small particles are located within the tilted attraction cone of the large colloid. Seldomly, we detect trimers formed by a big particle and two small particles, with the two small particles located in the side favored by the applied field. Since the spheres are firmly adsorbed to the fluid interface the different species always remain confined in the fluid interface. The magnetic torque induced by the vertical component of the applied field is never strong enough to overcome interfacial forces and induce the out-of-plane motion of the magnetic particles.[46]

The sudden inversion of the vertical component of the external field turns the previously formed dimers into a repulsive configuration, which promotes the ejection of the particles along the direction imposed by the horizontal component of the field (see **Figure 1a**). Although both particles repel one another, the smaller ones are expelled at a larger velocity since they possess a smaller drag coefficient. The latter can then be captured by another big particle in a disposition with its attractive cone exposed (Figure 1a, and **Movie1** in SI). This mechanism can be used to transfer small particles between nearby big colloids. Emission only occurs when the field is tilted enough to promote the mutual repulsion after the sudden reversing of the field (γ > 54,7º, regions II-IVb in Figure S1). The dynamics of the travelling particle is ballistic just after being ejected. In absence of adjacent neighboring big particles, the expelled particles diffuse on the water-decane interface as soon as they are outside the range of the emitter's dipolar field. The capture of the smaller particles is favored by the proximity between emitter and collector, and is observed when the tilt angle γ is smaller than 79º (≈ asin (Δz/($R_B$ + $R_s$)) + 54.7º ) (zones

I-IVa in Figure S1). As shown in **Figure 1b**, the described mechanism can also be used to transport two small particles or permanent aggregates of small particles between two nearby big spheres.

*2.1.2. Transference of Different-Sized Particles in Semi-Diluted Monolayers*

In monolayers with a relatively high area fraction of big spheres, $\varphi_B$, the application of fields with tilt angles in the range $54.7° < \gamma < $ asin $(\Delta z/(R_B + R_s)) + 54.7°$ induces the dipolar repulsion among big particles — located in the same plane—, favoring their self-assembly in an ordered lattice with local distortions due to the presence of small particles and aggregates.[47] The planar symmetry and the averaged distance between two big particles in the lattice are mainly determined by the strength and tilt angle of the applied field, the surface number density and the stoichiometry, i.e., the number ratio of the two different particles.[47, 48] In this configuration, most of the small particles form dimers with the big particles, while the rest diffuse in the interstitial region. The inversion of $H_z$ induces small particles to travel between two nodes (large particles) of the lattice, and this process can be repetitively reversed by reinverting the vertical component of the applied field, as shown in **Figure 2a**. Here, the applied field was recurrently reversed between $\mu_0H_0 = 3,5$ mT, $\gamma = 66.4°$ and $\mu_0H_0 = 3,5$ mT, $\gamma = -66.4°$, and the small particles were repeatedly ejected, at velocities close to 10 $\mu$ms$^{-1}$, and captured by different neighboring big particles. The observed trajectories can be understood by inspecting the potential energy landscape generated by the lattice of large particles at the fluid interface when they are under the tilted magnetic field. The energy profile, resembling a tilted washboard potential,[49] is asymmetric along the X direction, with an attractive well on one side and a repulsive part on the other (see **Figure 2b**). Upon the sudden inversion of $H_z$ the energy landscape changes, and the mutual interaction between large and small particles instantaneously passes from attractive (repulsive) to repulsive (attractive).

**Figure 2c** shows the time of flight $t_{flight}$, the average time taken by the small particles to travel between two big particles, for a monolayer with $\varphi_B = 0.27$, low stoichiometry, i.e. proportion of small and big magnetic colloids, and different values of $H_0$ and γ. In the range 55º < γ < 70º, $t_{flight}$ remains approximately constant at a given $H_x$, decreasing as $H_x$ increases. In this range of γ, the potential generated by the lattice is monotonic between the emitters and the collectors, and small particles are driven by the lattice magnetic potential (see Figure 2b). For low values of $H_x$, such magnetic potential is weak in the interstitial regions and the emitted particle diffuses before being captured by the collector. The lower $H_x$ is the more important diffusion is over ballistic motion, and $t_{flight}$ grows accordingly. For γ > 70º, $t_{flight}$ grows exponentially until γ ~ 79º, where no transit is observed. In the range 70º < γ < 79º, there is a potential energy barrier $U_B$ that colloids need to overcome to complete the transit towards the minimum energy configuration at contact. $U_B$ grows monotonically with the tilt angle, resulting in an exponential dependence of $t_{flight}$ on γ, according to an Arrhenius-Kramers law (see **Figure S2** in SI).[50, 51] For γ > 79º, the configuration of minimum energy occurs with the small colloid located at the interstitial region and no transit is observed. The increase in γ also induces the transition from a centered rectangular to a hexagonal lattice, both distorted by the presence of small aggregates.[47, 48, 52] This transition has been quantified in **Figure S3**.

**2.2 Transport of Confined Particles on the Self-Assembled Lattice**

*2.2.1 Transport Along any of the Crystallographic Directions*

To achieve net transport of the small magnetic particles on the lattice formed by large colloids we combine the periodic inversion of the vertical component $H_z$ with the precession of the field around the axis perpendicular to the interface:

$$\boldsymbol{H}(t) = \boldsymbol{H}_r(t, f_r) + \boldsymbol{H}_z(t, f_l) \tag{1}$$

with $H_r(t; f_r) = H_0(\sin\gamma \cos 2\pi f_r t, \sin\gamma \sin 2\pi f_r t, 0)$ and $H_z(t; f_I) = H_0(0, 0, \cos\gamma\, \text{sgn}(\sin \pi f_I t))$. Here, $f_I$ is the frequency of inversion. Under such actuation, the lattice of big colloidal particles is preserved, the dimers composed of small and large colloids are forced to rotate on the plane of the interface, while the small colloidal particles are ejected as soon as $H_z$ inverts its direction.

If the field inversion occurs at a frequency $f_I = 2f_r$, then the small particles spin around the collectors during half period before jumping to one of the adjacent particles in the lattice, along any of the crystallographic directions. In the rest of the article, unless stated otherwise, we will impose this restriction and focus on systems with low surface density of small particles. In **Figure 3a** (**Movie2** in SI) we show the trajectories of small particles on a hexagonal lattice with $\varphi_B = 0.27$ and $\gamma = 67°$. Under these conditions, the motion of the small particles is almost deterministic. For larger tilt angles, the energy profile presents a potential barrier and the emitted spheres diffuse in the interstitial area before being collected by neighboring big particles, randomizing the direction of transport (see **Figure 3b** and **Movie3** in SI, for the case with $\gamma = 78.7°$). Besides, particles can be transported on lattices with rectangular geometries by applying elliptically polarized precessing fields. If the rotating component of the precessing field is redefined as $H_r(t; f_r) = (H_x \sin\gamma \cos 2\pi f_r t, H_y \sin\gamma \sin 2\pi f_r t, 0)$, with $H_x > H_y$ and $a\tan\left(H_z/H_y\right) > 55°$, then the lattice constant along the X axis decreases while that along the Y direction increases to form a more centered rectangular structure.[48] In this configuration, the shorter cell edge is aligned along the X axis, as shown in **Figure 3c**.

We performed numerical simulations to investigate the dynamics displayed by small particles when the binary system of magnetic particles is under the actuation of the external field given by Equation 1, in conditions with $f_r = 2f_I$. The speed of the colloids $\langle v \rangle$, averaged over several transferences, exhibits a marked dependence on the tilt angle $\gamma$ with three distinct regimes, as shown in **Figure 3d**. For tilt angles 55º < $\gamma$ < 70º, the potential between the emitter and the collector nodes of the lattice is monotonic, the particles average velocity depends only on the frequency of the precessing field and the properties of the underlying lattice, being insensitive to the tilt angle. For 70º < $\gamma$ < 79º, the energy profile exhibits a potential barrier and the small magnetic colloids diffuse before being collected by the big particles. The size of the barrier increases with $\gamma$, causing a rapid decrease of the average particle velocity. For $\gamma$ > 79º, the interaction with the nodes of the underlying lattice is purely repulsive and the small particles are unable to leave the interstitial regions of the lattice.

For a given tilt angle $\gamma$ and precessing frequency, the dynamics of the small particles is also determined by the strength of the applied field. In **Figure 3e**, we show the results from numerical simulations for the time dependence of the mean square displacement (MSD) displayed by the small particles, for different magnitudes of the intensity of the dipolar-dipolar interaction, as measured by the dimensionless parameter $\Gamma = \mu_0 \chi^2 H^2 /(2\pi(R_B + R_s)^3 k_B T)$. Here, a loaded interface with $\varphi_B = 0.2$ is actuated by different fields with $\gamma = 67º$ and $f_r = 0.1$ Hz. Our results show that at short time scales the dynamics of the particles is diffusive. For cases with $\Gamma > 1$, however, we observe a superdiffusive regime at intermediate time scales with a crossover to a ballistic regime, at larger times. The crossover occurs at a characteristic time that increases with decreasing $\Gamma$. Note that we detect directed transport even for cases in which the big magnetic particles do not form a well-ordered 2D lattice (cases with $\Gamma < 100$).

The transport of small colloids is driven by the time-dependent precessing/inverting field, so their speed strongly depends on the frequency of actuation $f_I = 2f_r$. In **Figure 4a** we show the dependence of the average speed $\langle v \rangle$ measured as a function of $f_r$ for two different lattices with $\varphi_B$ = 0.30 or 0.13, in a case with low surface density of small particles ($\varphi_s$ < 0.05) to prevent mutual interactions. The inset images show the trajectories of small particles as they are driven by the external field. To ascertain the origin of the different dynamical regimes we performed numerical simulations of the system for different $\varphi_B$ and field frequencies, with good agreement with experiment (see **Figure 4b**). At low frequencies, the small colloids are synchronously driven by the applied field and the colloids average speed is approximately proportional to the field frequency. At a critical frequency $f_r^*(\varphi_B)$, there is a transition from the synchronous to an asynchronous mode, where the average particle velocity decreases with increasing actuation frequencies. The inception of this dynamical regime occurs when the period of field precession is comparable to the particles time of flight between emitter and collector lattice nodes. At these frequencies, the travelling particles reach the vicinity of the collectors when the induced dipoles are aligned in non-attractive configurations and both, the fraction of collected particles in a cycle and the average velocity, decrease. $f_r^*(\varphi_B)$ is inversely proportional to the time of flight of the colloidal particle, which increases with decreasing $\varphi_B$ (see **Figure 4S** in SI). At high frequency, the dimers rotate asynchronously with the external field,[53] and the expelled particles mainly feel the averaged attractive potential generated by the emitter particle.[54] As a result, the expelled particles almost immediately return to the emitter and the interfacial transport is blocked.

*2.2.2 Collective Transport*

The transport of the small magnetic particles through the 2D lattice is also strongly affected by collective effects and mutual interactions between traveling particles. In **Figure 4c**, we show numerical results for the flux of traveling particles as a function of stoichiometry of systems with different $\varphi_B$. The approximately linear dependence of the flux with $n_s/n_B$ at low stoichiometries soon acquires a concave behavior at intermediate concentrations ($n_s/n_B \sim 0.5$), reaching a saturated transport for $n_s/n_B \sim 1.0$ where the flow remains insensitive to the stoichiometry. The presence of small particles induces distortions on the underlying self-assembled lattice, affecting their transport. In addition, the competition for the collector nodes hinders the transfer of a larger number of traveling particles in concentrated systems. Furthermore, as the number of mobile particles increases the recoil of the underlying lattice increases too (inset in Figure 4c), reducing the velocity of the transported colloids with respect to the observational frame of reference. In **Figure 4d** we show the average velocity of the big and small particles, in a monolayer with an overall surface density of 0.1, as a function of $n_s/(n_{B+}\ n_s)$. The numerical results show that when the number of small particles becomes much larger than the number of big particles, the latter are transported in the opposite direction on self-assembled lattices composed by the smaller ones (Figure 4d). This inverted scenario is confirmed by the experiments (inset in Figure 4d and **Movie4** in SI).

*2.2.3 Monitored Transport*

The trajectory of the small particles is determined not only by the spatial distribution of the bigger particles or the tilt angle and strength of the applied, but also by the time pattern of the inversion of the vertical component. By monitoring the inversion times, the smaller particles can be driven through the net of bigger particles in any planned path. The maneuverability of the mechanism is illustrated in **Figure 5**, **Movie5** in SI, where a small particle is dynamically guided, following a preassigned closed curve around the frame of the microscope image. As shown in the movie, small aggregates in the net also rotate with the precessing field and can also be used as emitters and collectors. The launch direction is decided at any time, by inverting the vertical component of the precessing field in the right moment when the precessing dimmer points towards the desired collector.

## 3. Conclusion

Active or actuated colloids confined at a fluid-fluid interface are accessible model systems for studying collective transport in 2D or quasy-2D, at low Reynolds number, whether in the absence or presence of obstacles or pinning sites.[11] They show specific structures and dynamics not observable in the bulk of liquids.[20, 55] The strategy of transport introduced in this work, easily implemented but unique to strongly confined suspensions, takes advantage of the higher affinity of the particles to one of the mediums and the anisotropic character of the magnetic dipolar interaction to transport colloids or molecules onto fluid interfaces, through loosely-packed self-assembled lattices. In contrast to other experimental realizations, which require of sophisticated prefabricated substrates or optical ratchets,[6, 8, 56] here the active control over colloids is dynamically achieved thanks to the driven potential generated by the adsorbed particles themselves. The monitored actuation over the dynamic self-assembled structures allows for the reversible control of both, the symmetry/structure of the assembly and the transport mechanisms. Under the sudden reorientation of the applied field, the interactions between the different-sized particles are tuned to be either attractive or repulsive, enabling the rapid switching of colloids between two magnetic minima. For sufficiently high density, more complex translating potentials, generated by precessing/switching magnetic fields, are used to drive the particles through the array.

If the field is inverted with respect to the confining plane at a frequency double the precessing frequency, then the small particles spin around the collectors during half a complete rotation before jumping to one of the adjacent particles in the lattice. Under this condition, the speed of the colloids exhibits a marked dependence on the surface density as well as the tilt angle γ, frequency, and magnitude of the precessing/flashing field. We show that there exists a transition from the synchronous to an asynchronous mode of particle transport when the field rotation period is comparable to the particles time of flight between two adjacent lattice nodes. While the average velocity of the particles increases linearly with frequency in the synchronous mode at low frequencies, it decreases for frequencies above the transition frequency to the asynchronous mode.

The tilt angle of the external field with the interface also influences the dynamics of the travelling particles through the self-assembled lattice. We show that the average speed of the travelling particles is mostly insensitive to the tilt angle for 55º < γ < 70º, since in that regime the dipolar potential between the emitter and the collector nodes of the lattice is monotonic. As the tilt angle increases, the interaction potential develops an increasing energy barrier, causing a rapid decrease of the average particle velocity. Hence, the dynamics of the travelling colloids presents a crossover between a diffusive dynamics at short times and a ballistic regime at times much larger than the inversion time. The crossover time increases with increasing the tilt angle and/or decreasing the field strength.

The transport of the small magnetic particles through the 2D lattice is also strongly affected by collective effects. The increase in the number of small particles distorts the underlying self-assembled lattice and enhances the competition for collector nodes. As the number of small particles increases, the recoil of the underlying lattice increases, reducing the velocity of the small colloids with respect to the observational frame of reference, and promoting bidirectional transport.[57] A potential technical application of this system is based on the capability to induce the segregation of colloidal mixtures.[58] Finally, we also demonstrate that particles can be steered in any 2D-planned trajectory when the inversion of the precessing field is controlled on the fly.

These binary dispersions of magnetizable colloids, asymmetrically adsorbed onto a fluid interface, can be thought as a microscopically thin magnetic layered material,[59] with intralayer and interlayer dipolar interactions, as model for two-dimensional magnetic crystals,[60, 61] stacked semiconductor bilayers and strongly correlated dipolar Bose gases in two dimensions.[62-64] These smart systems, effectively switched or controlled by magnetic fields and free carrier doping, could also be realized by adsorbing nanoparticles or binary dispersions of equally sized polarizable particles with different wettability. They can be used for the manipulation and the transport of cargos and therapeutic payloads on the surface of liquids,[65] as convenient platforms for the development of optical switches or display technologies,[66] or the design of new functional self-assembled structures.[12, 35]

## 4. Experimental Section

*Experiments:* Superparamagnetic spherical microparticles with average sizes of $2R_B$ = 2.8 $\mu m$ and $2R_s$ = 1.0 $\mu m$ (Dynabeads M-270 and Dynabeads Myone, Invitrogen, respectively) were used in the experiments. The commercial solutions were washed twice in Milli-Q water to dilute the surfactant that may have been employed during the synthesis. Hence, particles were strongly confined on a fluid interface between two immiscible fluids. The bottom liquid was a 5.0 mm deep solution of NaCl 5 mM in water, with density $\rho$ = 1,08 kg m$^{-3}$ at 25 ºC, whereas decane from Dow Corning Corporation, with density $\rho$ =950 kg m$^{-3}$, viscosity $\eta$ =0.2×10$^{-4}$ m$^2$ s$^{-1}$ was used as a top liquid (3 mm deep). Particles, injected with a Pasteur pipette into the bulk of the bottom layer, are led to the interface via the application of a magnetic field gradient. Finally, the drift motion generated by convective effects is reduced by confining the loaded interface within a glass ring of 4.0 mm in diameter. According to the SEM images of the magnetic particles obtained with the Gel Trapping Technique (GET), the adsorbed particles are mostly immersed in the aqueous phase, with a small cap peaking above the interface (see inset in Figure 1a and Reference [20]). Therefore, they do not display structured monolayers of particles due to long-ranged dipole-dipole electrostatic repulsions,[67] neither aggregates due to capillary attractive interactions, even at relatively high surface coverages.[68, 69] By adsorbing different sized particles, we deliberately generate two parallel planes of induced dipoles separated by a vertical distance $\Delta z = R_B - R_s = 0.9\ \mu m$.

Same results are reproduced by adsorbing particles onto a water/air interface. Here, the dilute aqueous suspension of big and small particles is mixed with an aqueous solution of a cationic surfactant, typically 3 μM of dodecyltrimethylammonium bromide (DTAB). The mixed suspension is poured in a glass cylindric tube, 7.0 mm in height and 5.0 mm in diameter, previously glued to a glass slide. Afterwards, the whole system is inverted. In this configuration, the pendant suspension remains contained by vessel thanks to the capillary forces, and the adsorption of the settling particles is favored the presence of the cationic surfactant. After 60 minutes, the small vessel is brought to its initial position, where the aqueous suspension is the bottom phase and the monolayer of adsorbed particles can be easily visualized with a straight microscope (the method is described in detail in References [70, 71]). The main advantage of this second method is that it does not require of the application of magnetic field gradients, which may promote the formation of permanent aggregates. The formation of small irreversible aggregates is also favored by the presence of the salt or the cationic surfactant in the aqueous dispersion, both used for facilitating the adsorption of the particles on the fluid interface.

The colloidal monolayer is energized by a time-dependent external field, which drives the system out of equilibrium and controls the emerging structures. The experimental set-up was custom built and originally described in Reference [20]. The loaded interface is placed in the middle of two pairs of orthogonal coils, aligned along the X and Y axes, while a third coil is placed under the sample, aligned along the optical axis of the microscope. The coils aligned along the X and Y axes are connected to a power amplifier (Kepco BOP 2X20-10D) commanded by a waveform generator (National Instruments 9269). The coil aligned along the optical axis is connected to a DC power supply (Tenma, 72-2805). The resultant field is constant and applied along the XZ plane, or precesses around the Z axis at an angular frequency $\omega_r$. Finally, the adsorbed monolayer is visualized with an upright optical microscope (BH2, Olympus) connected to a CCD camera (EO1312M, Edmund) equipped with a 40 × 0.65 NA and a 20 × 0.40 NA objective.

*Numerical Simulations:* Our understanding of the experiments is complemented by a minimalistic numerical model based on Brownian dynamics simulations of a point-dipole model of the vertically confined paramagnetic colloids, where each particle is described by a dipole moment $m = \mu_0 \chi (H_{ext} + H_{ind})$ located at its center. Here, $H_{ext}$ is the external applied field and Hind is the field induced by the rest of dipoles in the system, which is calculated in a self-consistent manner. The colloids interact via magnetic dipole-dipole and repulsive steric interactions which provide them with extension. The magnetic dipole-dipole interaction colloids are much weaker than the interfacial forces which keep the colloids trapped at the water-decane (or water-air) interface. Consequently, in our model we have considered that the vertical positions of each type of colloid, determined by its size and the wetting angle, are fixed throughout the simulation.

**Supporting Information**
Supporting Information is available from the Wiley Online Library or from the author.

**Acknowledgements**
This work was funded by Horizon 2020 program through 766972-FET-OPEN-NANOPHLOW, and by MINECO under the grant CTQ2016-78895-R. C.C. is grateful to I. Pagonabarraga for fruitful discussions. F.M.-P. acknowledges support from MINECO (Grant Nº. RYC-2015-18495) and UCM/SANTANDER 2019 (PR87/19).

References

[1]     M. Pelton, K. Ladavac, D. G. Grier, Physical Review E 2004, 70, 031108.
[2]     U. Choudhury, A. V. Straube, P. Fischer, J. G. Gibbs, F. Höfling, New J. Phys. 2017, 19, 125010.
[3]     J. Repp, G. Meyer, F. E. Olsson, M. Persson, Science 2004, 305, 493.
[4]     R. Hoffmann-Vogel, Applied Physics Reviews 2017, 4, 031302.
[5]     S. Coh, W. Gannett, A. Zettl, M. L. Cohen, S. G. Louie, Physical review letters 2013, 110, 185901.
[6]     C. Drexler, S. A. Tarasenko, P. Olbrich, J. Karch, M. Hirmer, F. Müller, M. Gmitra, J. Fabian, R. Yakimova, S. Lara-Avila, S. Kubatkin, M. Wang, R. Vajtai, P. M. Ajayan, J. Kono, S. D. Ganichev, Nature Nanotechnology 2013, 8, 104.
[7]     K. Jacobson, P. Liu, B. C. Lagerholm, Cell 2019, 177, 806.
[8]     P. Tierno, F. Sagués, T. H. Johansen, T. M. Fischer, Physical Chemistry Chemical Physics 2009, 11, 9615.
[9]     X.-g. Ma, P.-Y. Lai, B. J. Ackerson, P. Tong, Soft Matter 2015, 11, 1182.
[10]    X. Cao, E. Panizon, A. Vanossi, N. Manini, C. Bechinger, Nature Physics 2019, 15, 776.
[11]    K. Dietrich, G. Volpe, M. N. Sulaiman, D. Renggli, I. Buttinoni, L. Isa, Physical review letters 2018, 120, 268004.
[12]    A. Snezhko, I. S. Aranson, Nat Mater 2011, 10, 698.
[13]    H. Ebata, M. Sano, Scientific Reports 2015, 5, 8546.
[14]    G. Grosjean, M. Hubert, G. Lagubeau, N. Vandewalle, Physical Review E 2016, 94, 021101.
[15]    E. Bormashenko, Y. Bormashenko, R. Grynyov, H. Aharoni, G. Whyman, B. P. Binks, The Journal of Physical Chemistry C 2015, 119, 9910.
[16]    F. Martinez-Pedrero, A. Ortiz-Ambriz, I. Pagonabarraga, P. Tierno, Physical review letters 2015, 115, 138301.
[17]    G. Grosjean, M. Hubert, Y. Collard, S. Pillitteri, N. Vandewalle, Eur. Phys. J. E 2018, 41, 137.
[18]    T. Li, A. Zhang, G. Shao, M. Wei, B. Guo, G. Zhang, L. Li, W. Wang, Advanced Functional Materials 2018, 28, 1706066.
[19]    J. J. Giner-Casares, J. Reguera, Nanoscale 2016, 8, 16589.
[20]    F. Martínez-Pedrero, F. Ortega, J. Codina, C. Calero, R. G. Rubio, J. Colloid Interface Sci. 2020, 560, 388.
[21]    J. Engström, C. J. Brett, V. Körstgens, P. Müller-Buschbaum, W. Ohm, E. Malmström, S. V. Roth, Advanced Functional Materials, n/a, 1907720.
[22]    W. Fei, Y. Gu, K. J. M. Bishop, Current Opinion in Colloid & Interface Science 2017, 32, 57.
[23]    F. Berger, C. Keller, Melanie J. I. Müller, S. Klumpp, R. Lipowsky, Biochemical Society Transactions 2011, 39, 1211.
[24]    A. Bricard, J.-B. Caussin, N. Desreumaux, O. Dauchot, D. Bartolo, Nature 2013, 503, 95.
[25]    F. Martinez-Pedrero, E. Navarro-Argemí, A. Ortiz-Ambriz, I. Pagonabarraga, P. Tierno, Science Advances 2018, 4 eaap9379.
[26]    H. M.-C. a. A. O.-A. a. A. V. a. P. Tierno, arXiv e-prints 2019, 1911.01698.

[27] F. Meng, A. Ortiz-Ambriz, H. Massana-Cid, A. Vilfan, R. Golestanian, P. Tierno, Physical Review Research 2020, 2, 012025.
[28] R. Hartmann, P. K. Singh, P. Pearce, R. Mok, B. Song, F. Díaz-Pascual, J. Dunkel, K. Drescher, Nature Physics 2019, 15, 251.
[29] A. E. Patteson, A. Gopinath, P. E. Arratia, Nat. Commun. 2018, 9, 5373.
[30] D. Needleman, Z. Dogic, Nature Reviews Materials 2017, 2, 17048.
[31] K. Doxzen, S. R. K. Vedula, M. C. Leong, H. Hirata, N. S. Gov, A. J. Kabla, B. Ladoux, C. T. Lim, Integrative Biology 2013, 5, 1026.
[32] H.-W. Huang, F. E. Uslu, P. Katsamba, E. Lauga, M. S. Sakar, B. J. Nelson, Science Advances 2019, 5, eaau1532.
[33] I. H. Riedel, K. Kruse, J. Howard, Science 2005, 309, 300.
[34] S. Kriegman, D. Blackiston, M. Levin, J. Bongard, Proceedings of the National Academy of Sciences 2020, 201910837.
[35] B. Wang, K. F. Chan, J. Yu, Q. Wang, L. Yang, P. W. Y. Chiu, L. Zhang, Advanced Functional Materials 2018, 28, 1705701.
[36] M. Driscoll, B. Delmotte, Current Opinion in Colloid & Interface Science 2019, 40, 42.
[37] T. Vicsek, A. Czirók, E. Ben-Jacob, I. Cohen, O. Shochet, Physical review letters 1995, 75, 1226.
[38] M. S. Davies Wykes, J. Palacci, T. Adachi, L. Ristroph, X. Zhong, M. D. Ward, J. Zhang, M. J. Shelley, Soft Matter 2016, 12, 4584.
[39] L. Berthier, E. Flenner, G. Szamel, The Journal of Chemical Physics 2019, 150, 200901.
[40] C. Reichhardt, C. J. Olson Reichhardt, Physical Review E 2014, 90, 012701.
[41] C. Reichhardt, C. J. O. Reichhardt, Soft Matter 2014, 10, 2932.
[42] S. Garcia, E. Hannezo, J. Elgeti, J.-F. Joanny, P. Silberzan, N. S. Gov, Proc. Natl. Acad. Sci. U. S. A. 2015, 112, 15314.
[43] R. L. Stoop, A. V. Straube, P. Tierno, Nano letters 2019, 19, 433.
[44] J. Palacci, S. Sacanna, A. P. Steinberg, D. J. Pine, P. M. Chaikin, Science 2013, 339, 936.
[45] B. M. Mognetti, A. Šarić, S. Angioletti-Uberti, A. Cacciuto, C. Valeriani, D. Frenkel, Physical review letters 2013, 111, 245702.
[46] S. Cappelli, Vol. Phd Thesis 1 (Research TU/e / Graduation TU/e), Technische Universiteit Eindhoven, Eindhoven 2016/9/28.
[47] L. J. Bonales, F. Martínez-Pedrero, M. A. Rubio, R. G. Rubio, F. Ortega, Langmuir : the ACS journal of surfaces and colloids 2012, 28, 16555.
[48] W. Wen, L. Zhang, P. Sheng, Physical review letters 2000, 85, 5464.
[49] C. Cheng, M. Cirillo, G. Salina, N. Grønbech-Jensen, Physical Review E 2018, 98, 012140.
[50] H. A. Kramers, Physica 1940, 7, 284.
[51] N. Berglund, arXiv e-prints 2011, 1106.5799
[52] U. Gasser, C. Eisenmann, G. Maret, P. Keim, ChemPhysChem 2010, 11, 963.
[53] G. Helgesen, P. Pieranski, A. T. Skjeltorp, Physical review letters 1990, 64, 1425.
[54] N. Osterman, I. Poberaj, J. Dobnikar, D. Frenkel, P. Ziherl, D. Babić, Physical review letters 2009, 103, 228301.
[55] B. P. Binks, Langmuir : the ACS journal of surfaces and colloids 2017, 33, 6947.
[56] A. V. Arzola, M. Villasante-Barahona, K. Volke-Sepúlveda, P. Jákl, P. Zemánek, Physical review letters 2017, 118, 138002.
[57] F. Martinez-Pedrero, H. Massana-Cid, T. Ziegler, T. H. Johansen, A. V. Straube, P. Tierno, Physical Chemistry Chemical Physics 2016, 18, 26353.

[58]     M. P. MacDonald, G. C. Spalding, K. Dholakia, Nature 2003, 426, 421.
[59]     K. F. Mak, J. Shan, D. C. Ralph, Nature Reviews Physics 2019, 1, 646.
[60]     C. Gong, X. Zhang, Science 2019, 363, eaav4450.
[61]     L. Huang, Z. Hu, H. Jin, J. Wu, K. Liu, Z. Xu, J. Wan, H. Zhou, J. Duan, B. Hu, J. Zhou, Advanced Functional Materials, n/a, 1908486.
[62]     C. Hubert, Y. Baruchi, Y. Mazuz-Harpaz, K. Cohen, K. Biermann, M. Lemeshko, K. West, L. Pfeiffer, R. Rapaport, P. Santos, Physical Review X 2019, 9, 021026.
[63]     D. Hufnagl, R. E. Zillich, Physical Review A 2013, 87, 033624.
[64]     A. Macia, D. Hufnagl, F. Mazzanti, J. Boronat, R. E. Zillich, Physical review letters 2012, 109, 235307.
[65]     M. Luo, Y. Feng, T. Wang, J. Guan, Advanced Functional Materials 2018, 28, 1706100.
[66]     A. F. Demirörs, P. J. Beltramo, H. R. Vutukuri, ACS Applied Materials & Interfaces 2017, 9, 17238.
[67]     P. Pieranski, Physical review letters 1980, 45, 569.
[68]     S. Cappelli, A. M. de Jong, J. Baudry, M. W. J. Prins, Langmuir : the ACS journal of surfaces and colloids 2017, 33, 696.
[69]     B. J. Park, E. M. Furst, Soft Matter 2011, 7, 7676.
[70]     M. Anyfantakis, J. Vialetto, A. Best, G. K. Auernhammer, H.-J. Butt, B. P. Binks, D. Baigl, Langmuir : the ACS journal of surfaces and colloids 2018, 34, 15526.
[71]     A. Maestro, E. Guzmán, E. Santini, F. Ravera, L. Liggieri, F. Ortega, R. G. Rubio, Soft Matter 2012, 8, 837.

**Figure 1.** a) The scheme and image sequence describe the strategy used to transfer small particles between nearby big spheres. After flipping the vertical component of the tilted field, a small particle is repelled by its neighbor in the dimmer. Once far away from the emitter, the expelled particle diffuses on the water-decane interface, and subsequently is captured by a neighboring big particle disposed in an attractive configuration, with the center-to-center line within the attractive cone of the dipole-dipole interaction (red triangles in the scheme). The SEM image in the inset, obtained with GTT, shows how the adsorbed particles are mostly immersed in the aqueous phase, with a small cap peaking above the interface (see Reference [20]). b) The described mechanism is also used to transport two small particles (i and ii) or permanent aggregates of small particles (iii) between two nearby big spheres. In a) and b), the field is changed from µ0H0 = 3,5 mT, γ = 66.4º to µ0H0 = 3,5 mT, γ = -66.4º.

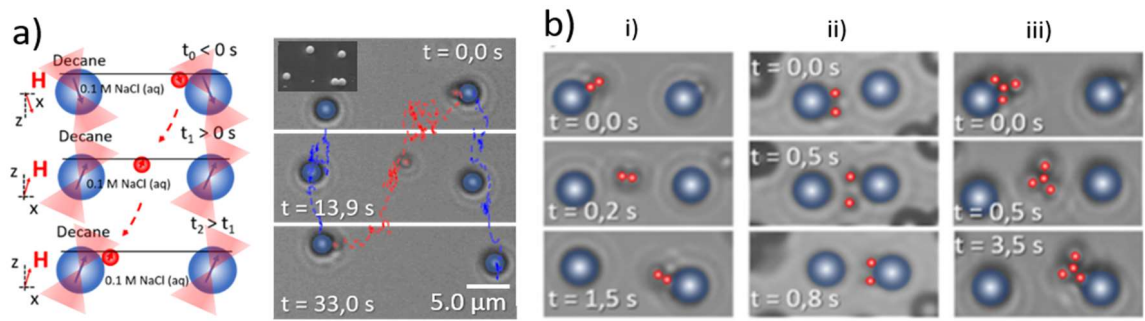

**Figure 2.** a) The periodic flipping of the applied field between $\mu_0H_0 = 3,5$ mT, $\gamma = 66.4°$ and $\mu_0H_0 = 3,5$ mT, $\gamma = -66.4°$, forces the small particle to successively travel between two nodes of the lattice. In the previous experimental conditions, the particles are ballistically ejected at velocities close to 10 µms$^{-1}$, diffuse (or not) in the interstices and finally are captured by the neighboring big particles. b) Magnetostatic potential created by the big spheres at the fluid interface level, before (left) and after (right) inverting the vertical component of the applied field for $\mu_0H_0 = 2$mT and $\gamma = 67°$. Graphs at the bottom show the energy landscapes along the lowest-energy curve, the dashed line that connects two big particles in the lattice, for different $\gamma$ values. c) The average time taken by the small particles to travel between two big particles through the viscous media, $t_{flight}$, for a monolayer with $\varphi_B = 0.27$ and a minor number of small spheres. The horizontal component of the applied field is $\mu_0H_x = 2.85$ mT (■), 1.44 mT (●), 2.15 mT (▲), 2.86 mT (○), 3.57 mT (□). Errors are not included in the graph for the sake of clarity.

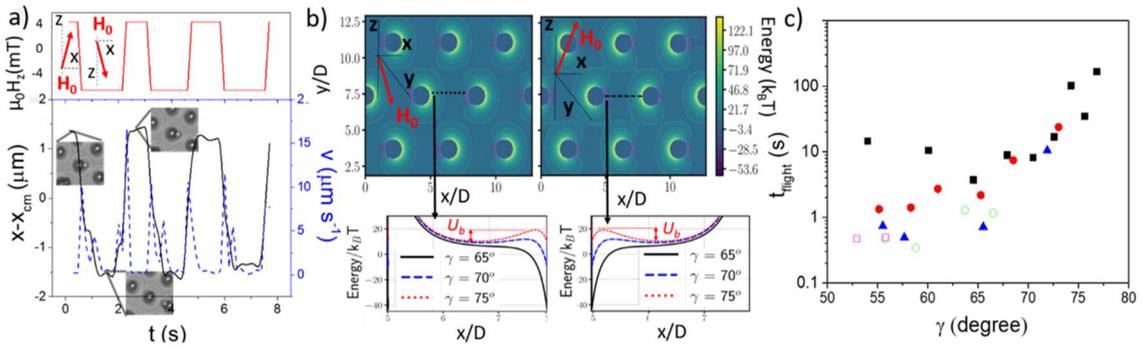

**Figure 3.** a) For $\varphi_B = 0.27$, $f_r = 0.1$ Hz and $\gamma = 67°$, $\mu_0 H_r = 1.0$ mT and $\mu_0 H_z = 2.4$ mT, the big particles form a hexagonal lattice while the big-small dimmers rotate. Small particles are driven through the lattice, in a practically deterministic 1D motion, by periodically inverting the vertical component of $H_z$, at a frequency $f_I = \omega_r/\pi = 2f_r$. b) For $\varphi_B = 0.27$, and applying a precessing magnetic field with $\mu_0 H_r = 0.5$ mT, $\mu_0 H_z = 2.5$ mT, $f_r = 0.1$ Hz and $\gamma = 79.5°$, the emitted spheres diffuse in the interstitial area before being collected by neighboring big particles. c) For $\varphi_B = 0.20$, and applying an elliptically polarized precessing magnetic field, with $\mu_0 H_x = 1.6$ mT, $\mu_0 H_y = 0.4$ mT, $\mu_0 H_z = 2.4$ mT and $f_r = 0.1$ Hz, the lattice constant along the x axis decreases while that along the y direction increases, and the transport is facilitated in the x direction. d) Average velocity per particle $<v>$ as a function of the tilt angle of the precessing field $\gamma$ as obtained from numerical simulations, with $\varphi_B = 0.2$, $f_r = 0.1$ Hz and $\mu_0 H_r = 2$ mT. Here, $v_0 = (\pi R_B^2/\varphi_B)^{1/2} f_r$ is a characteristic velocity for a given lattice, with $\varphi_B$ and frequency $f_r$. e) Time-dependence of MSD of the small particles for a system with $\varphi_B = 0.2$, actuated by fields giving different values of $\Gamma$, $\gamma = 67°$ and $f_r = 0.1$ Hz. Here, $l = (\pi R_B^2/\varphi_B)^{1/2}$ is a measure of the lattice spacing.

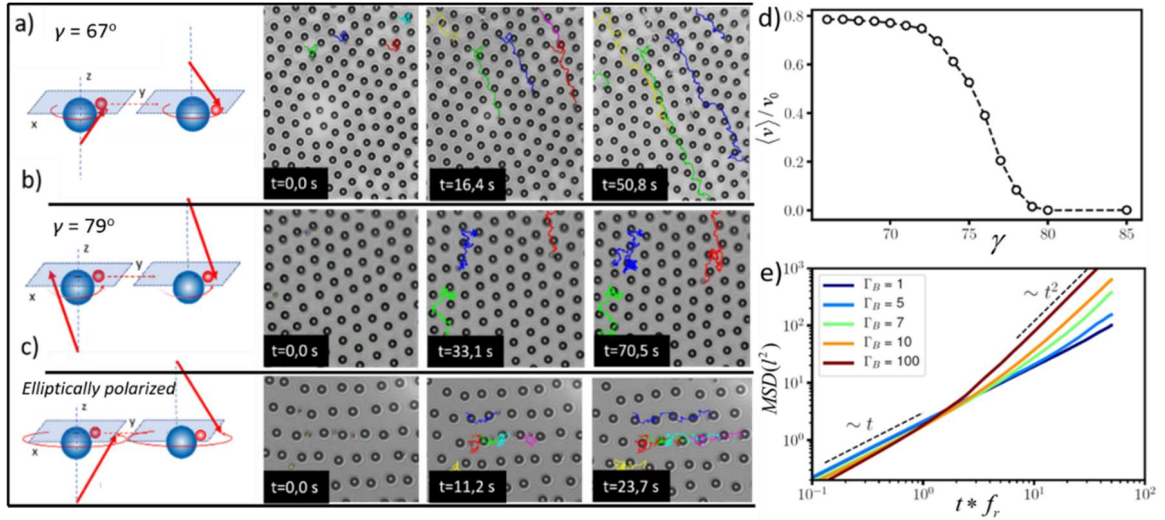

**Figure 4:** a) The average velocity <v> of the transported particles as a function of $f_r$, in two different monolayers with $\varphi_B = 0.30$ (○) and $\varphi_B = 0.13$ (■). The precessing and flashing magnetic field, with $\mu_0 H_r = 1.0$ mT, $\mu_0 H_z = 2.4$ mT and $\gamma = 67°$ is applied imposing $f_I = \omega_r/\pi = 2f_r$. In the inserted images $f_I = 2f_r = 0.2$ Hz, $\varphi_B = 0.13$ and $\varphi_s = 0.01$ (left) and $\varphi_B = 0.30$ and $\varphi_s = 0.02$ (right). b) Numerical results for the average speed of transported particles <v> as a function of $f_r$, for monolayers with different $\varphi_B$, $\mu_0 H_0 = 2.5$ mT and $\gamma = 67°$. c) Flux of particles as a function of the stoichiometry of the binary system for $\mu_0 H_r = 2$ mT, $f_r = 0.04$ Hz and $\gamma = 67°$. Inset: recoil average velocity of lattice as a function of stoichiometry for different $\varphi_B$. d) Average velocity <v> of small particles (○) and big particles (■) of systems with different fraction of small particles $n_S/(n_S + n_B)$ under the actuation of a magnetic field with $\mu_0 H_r = 2$ mT, $f_r = 0.2$ Hz and $\gamma = 67°$. Here, the overall surface density is 0.1.

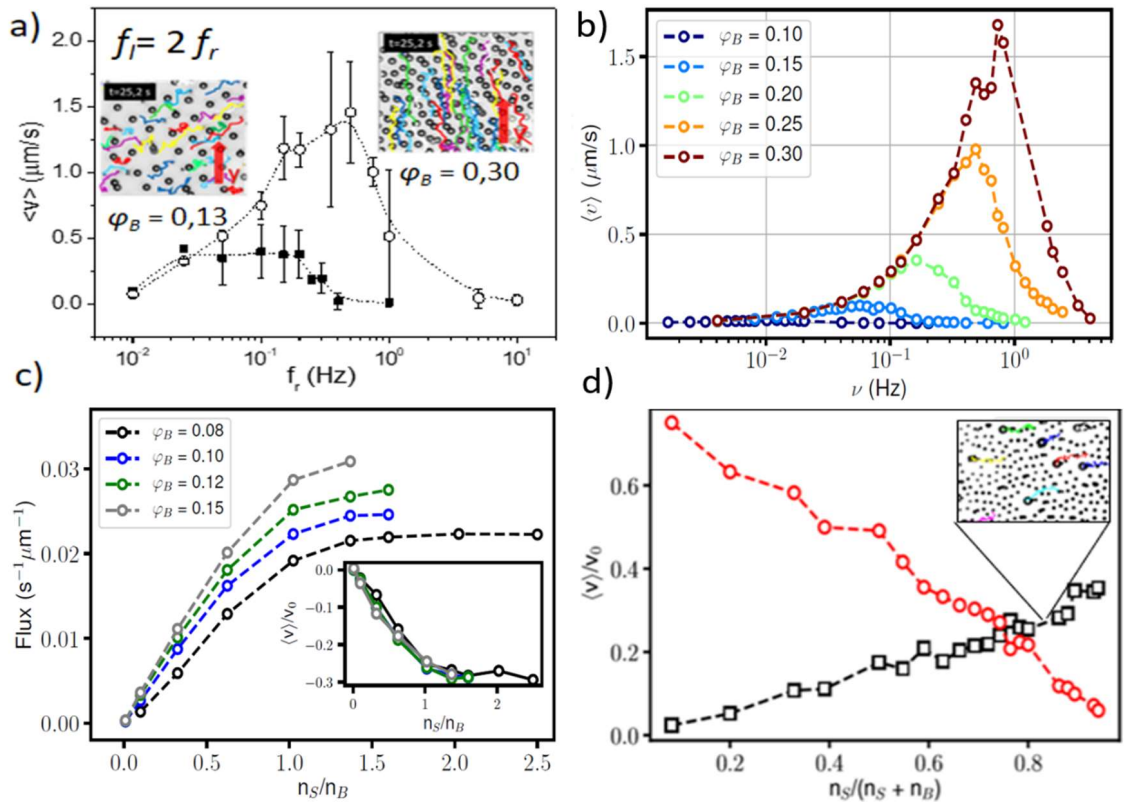

**Figure 5:** The sequence of images shows how a small particle is driven in a planned trajectory, through the net of bigger particles adsorbed on the water/decane interface, by monitoring the inversion times. The colors in the superimposed line and schematic have been included to identify how the recurrent inversion of the precessing field $H(t) = H_r(f_r) + H_z(f_i)$, with $\mu_0 H_r = 1.0$ mT, $\mu_0 H_z = 2.7$ mT, and $f_i = 0.1$ Hz, is used to steer the particles through the lattice.

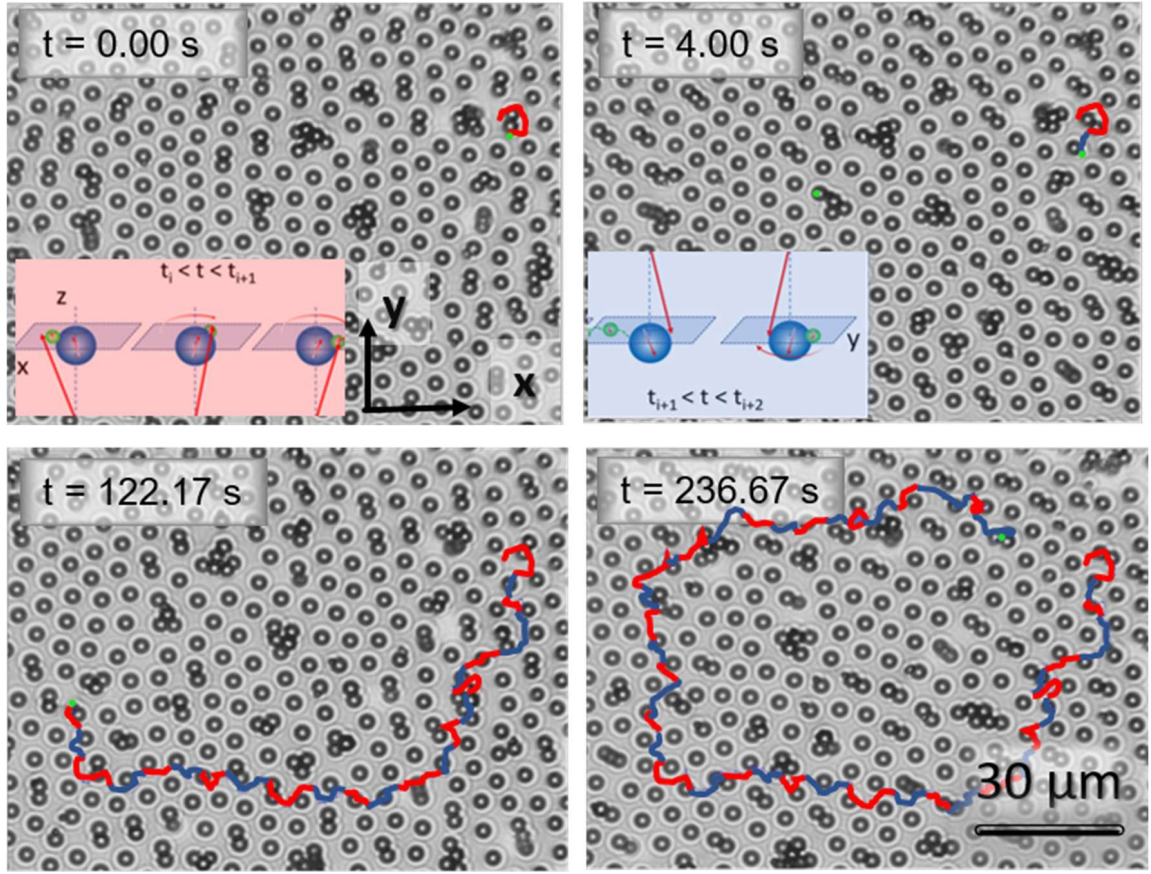

# Supporting Information

**Collective Transport of Magnetic Microparticles at a Fluid Interface through Dynamic Self-Assembled Lattices**

*Fernando Martínez-Pedrero[1,*], Francisco Ortega[1,2], Ramón G.Rubio[1,2], Carles Calero[3,4,*]*

**I. State Diagram**

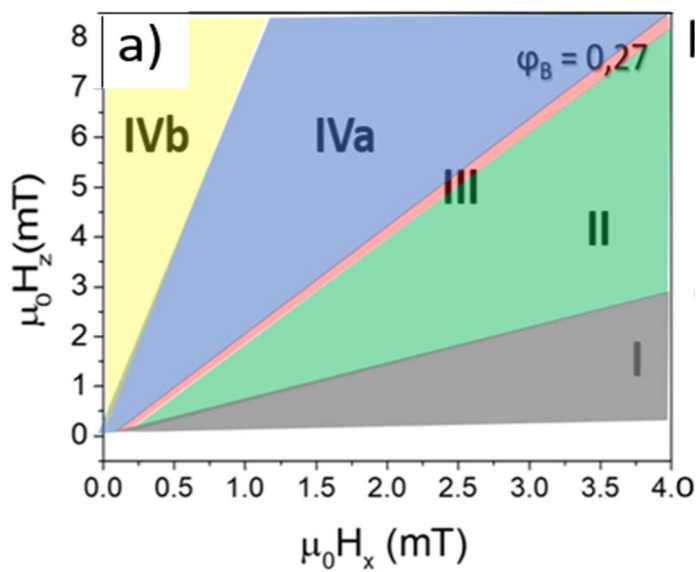

Figure S1: In relatively diluted binary monolayers, the application of a field in the X direction, parallel to the interface, induces the formation of linear aggregates composed by small and big particles and aligned along the X axis. The subsequent application of an extra field along the z axis, $H_z$, increases the angle between the induced dipoles and the confining plane, modulating the magnetic dipole-dipole interaction between the colloidal particles and promoting the repulsion between the particles, the fragmentation of the linear aggregates and the generation of a set of new states, determined by the $H_z/H_x$ ratio. At low $H_z$, zone I, the dipolar interaction between the particles still promotes the formation of binary chains oriented along the x axis. In zone II, the application of the vertical component tilts the induced moments along the direction of the field, causing the
26

rupture of the bonds formed between big and small particles, when the line connecting the particles and the applied field have slopes of opposite sign. In zone III, the binary monolayer slowly arrives to a configuration where monomers and binary chains coexist. As the applied field is further leaned towards the vertical direction, zone IV, all equal sized spheres repel each other, and the binary linear aggregates disassemble through an abrupt colloidal explosion. The system quickly turns into a monolayer of monomers, dimmers, constituted by big and small particles pointing along the X direction, and a small proportion of permanent aggregates. In monolayers with a relatively high surface number density of big spheres, $\varphi_B$, the dipolar repulsion induced via the application of a tilted field in zone IVa-b causes them to arrange on a partially ordered structure, locally distorted by the presence of both small aggregates and small particles. The planar symmetry and the averaged distance between two big particles in the net is mainly determined by the strength and tilt angle of the applied field, the surface number density and the stoichiometry (i.e., the number ratio of the two different particles). In the zone IV, the sudden inversion of the vertical component of the external field turns the previously formed dimers into a repulsive configuration. Emission only occurs when the field is tilted enough to promote the mutual repulsion after the sudden reversing of the field (zone IVb). The capture of the smaller particles is favored by the proximity between emitter and collector, and only happens when the tilt angle $\gamma = a\tan(H_z/H_x)$ is smaller than 79º (zone IVa).

## II. Energy barrier in the range 70º < γ < 79º

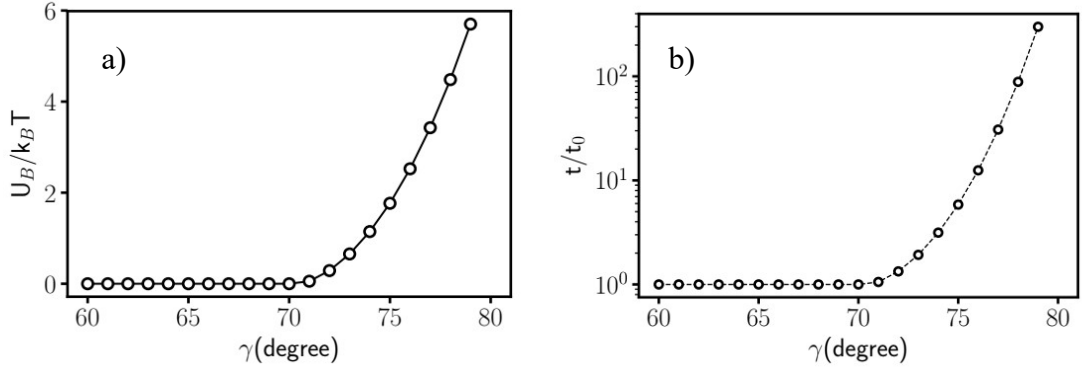

Figure S2: For tilt angles in the range 70º < γ < 79º, the dipolar potential exerted by the self-assembled lattice on the small particles exhibits an energy barrier $U_B$ along the path between the emitter and the collector. The height of the barrier increases monotonically with γ, see Figure S2a. The transit of the small particle from the emitter to the collector (the minimum energy configuration is at contact) needs to overcome such energy barrier. The time of transit for an overdamped brownian particle under these circumstances follows an Arrhenius-Kramers law, t ~ $\exp(U_B/k_BT)$ ~ $\exp(\gamma/k_BT)$,[50, 51] as shown in Figure S2b and in qualitative accordance with experiment (see Figure 2c in the main text).

## III. Distortion of lattice with tilt angle

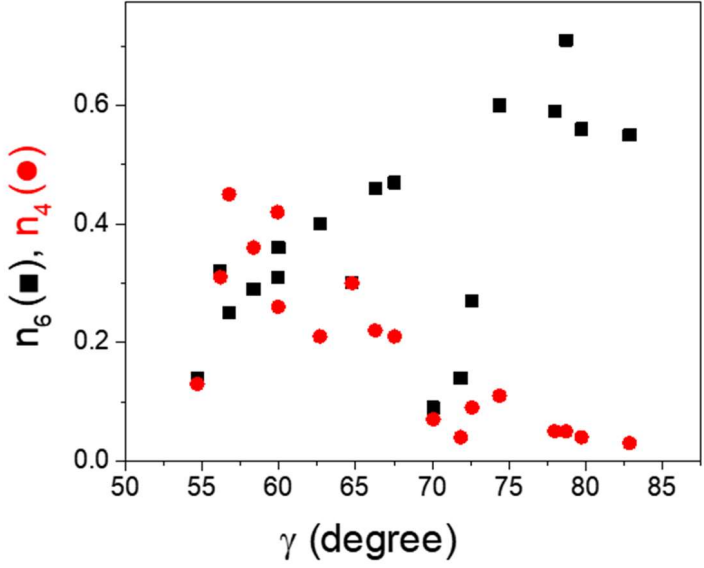

Figure S3: The increase in $\gamma$ induces the transition from a centered rectangular to a hexagonal lattice, both distorted by the presence of small aggregates. This transition has been quantified by calculating the local bond orientational order parameter, defined by $n_p \equiv \frac{1}{N}\left|\sum_{k=1}^{N} \varphi_{p,k}\right|$, where N is the total number of particles in the image and $\varphi_{p,k} \equiv \frac{1}{n}\sum_{j=1}^{n} e^{ip\theta_{kj}}$. Here, the sum is carried out over the $n$ nearest neighbors and $\theta_{kj}$ is the angle between the axis x and the bond joining neighbors $i$ and $j$. Delaunay triangulation underlying the Voronoi tessellation allows us to evaluate both, $n$ and the angle between the segments connecting them. Here, the applied field was $\vec{H}_0 = \vec{H}_x + \vec{H}_z$, with $\mu_0 H_z = 3.2$ mT, and $\mu_0 H_x = 2.85$ mT, 1.44 mT, 2.15 mT, 2.86 mT and 3.57 mT.

## IV. Transition frequency to asynchronous mode

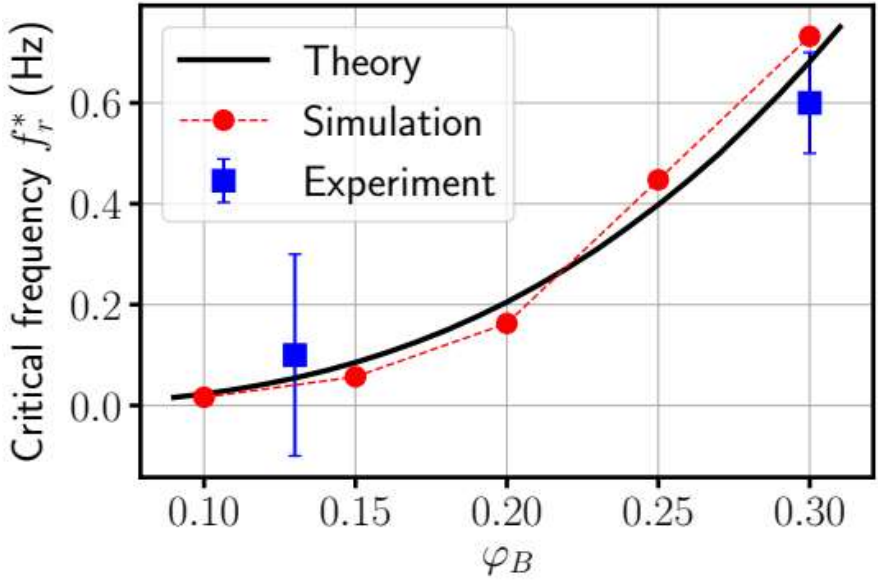

Figure S4: For external fields precessing with 55º < γ < 70º, the travelling particles are driven by the magnetic potential created by the underlying lattice, which monotonically decays with the distance between the emitter and the collector nodes. Being in the overdamped regime, the velocity of the particles is proportional to the dipolar force, v = F/(6πηR$_S$), and the time taken to transit between two consecutive nodes of the lattice t$_{flight}$ can be easily calculated theoretically for a given $\varphi_B$. The transition from the synchronous to the asynchronous mode occurs when the transit time is comparable to the precessing frequency, which defines a critical frequency $f_r^*(\varphi_B)$ = 1/t$_{flight}$. We can also assess the critical frequencies from the numerical simulation results, as those frequencies at which the velocity of the transported particles is maximum (see Figure 4b of the manuscript). In the same way, we measure the experimental critical frequency for two different values of $\varphi_B$, which are in good agreement with the values obtained from the simulations and theoretical predictions. The Figure shows the critical frequency $f_r^*$ as a function of $\varphi_B$, obtained from theoretical arguments, simulation results, and experiments.

**Movie1:** A small magnetic particle, 1.0 μm in size, is transmitted between two bigger particles, 2.8 μm in size, along the fluid interface. The transfer is triggered by the sudden inversion of the applied field, which is changed from $\mu_0 H_0 = 3,5$ mT, $\gamma = 66.4°$ to $\mu_0 H_0 = 3,5$ mT, $\gamma = -66.4°$. The movie is sped up ten times.

**Movie2:** Small particles 1.0 μm in size, are driven through a hexagonal lattice, formed by bigger particles, 2.8 μm in size, with $\varphi_B = 0.27$. The horizontal component of the precessing field, with $\mu_0 H_r = 1.0$ mT and $\mu_0 H_z = 2.4$ mT and $\gamma = 67.0°$, is periodically inverted at a frequency $f_I = 0.2 = 2f_r$.

**Movie3:** Small particles 1.0 μm in size, are driven through a hexagonal lattice, formed by bigger particles, 2.8 μm in size, with $\varphi_B = 0.27$. The horizontal component of the precessing field, with $\mu_0 H_r = 0.5$ mT and $\mu_0 H_z = 2.5$ mT and $\gamma = 79.5°$, is periodically inverted at a frequency $f_I = 0.2 = 2f_r$.

**Movie4:** Big particles 2.8 μm in size, are driven through a random lattice, formed by smaller particles, 1.0 μm in size, with $\varphi_B = 0.19$. The horizontal component of the precessing field, with $\mu_0 H_r = 1.0$ mT and $\mu_0 H_z = 2.4$ mT and $\gamma = 67.0°$, is periodically inverted at a frequency $f_I = 0.5 = 2f_r$. The movie is sped up ten times.

**Movie5:** A small particle 1.0 μm in size, is driven through a random lattice, formed by bigger particles, 2.8 μm in size, and small aggregates, following a preassigned closed curve around the frame of the microscope image. The inversion of the horizontal component of the precessing field, with $\mu_0 H_r = 1.0$ mT, $f_r = 0.1$ Hz and $\mu_0 H_z = 2.7$ mT, is monitored for steering the particles through the lattice. The movie is sped up ten times.